\documentclass{Interspeech}
\usepackage{amssymb,amsmath,epsfig,cite,url,multicol,multirow}
\usepackage{amsmath,graphicx}
\usepackage{xcolor}
\usepackage{hyperref}
\usepackage{fancyvrb}
\usepackage{fvextra} 
\usepackage{graphicx}
\usepackage{float}
\usepackage{comment}
\usepackage{float}
\usepackage{diagbox,balance}
\floatstyle{plaintop}
\restylefloat{table}
\usepackage{amssymb}
\usepackage{pifont}

\interspeechcameraready

\title{Better Pseudo-labeling with Multi-ASR Fusion and Error Correction by SpeechLLM}

\author{Jeena Prakash$^*$, Blessingh Kumar$^*$, Kadri Hacioglu, Bidisha Sharma, Sindhuja Gopalan,\\
Malolan Chetlur, Shankar Venkatesan, Andreas Stolcke\\
\\
Uniphore Systems, India \& USA}

\email{\{jeena, blessingh, kadri.hacioglu, bidisha, sindhuja, malolan.chetlur, shankar.venkatesan, andreas.stolcke\}uniphore.com}

\usepackage{comment}

\begin{document}

\maketitle
\def\thefootnote{*}\footnotetext{These authors contributed equally to this work}
\begin{abstract}
Automatic speech recognition (ASR) models rely on high-quality transcribed data for effective training. Generating pseudo-labels for large unlabeled audio datasets often relies on complex pipelines that combine multiple ASR outputs through multi-stage processing, leading to error propagation, information loss and disjoint optimization. We propose a unified multi-ASR prompt-driven framework using postprocessing by either textual or speech-based large language models (LLMs), replacing voting or other arbitration logic for reconciling the ensemble outputs. We perform a comparative study of multiple architectures with and without LLMs, showing significant improvements in transcription accuracy compared to traditional methods. Furthermore, we use the pseudo-labels generated by the various approaches to train semi-supervised ASR models for different datasets, again showing improved performance with textual and speechLLM transcriptions compared to baselines.
\end{abstract}

\vspace{-0.2cm}
\section{Introduction}
To achieve good generalization, an ASR model must be trained on diverse datasets that capture a wide range of accents, dialects, genres, and other speech patterns. Although semi-supervised and self-supervised learning methods have reduced the need for high-quality audio datasets for model training, supervised ASR models are still crucial for commercial applications in many domains.
Obtaining such datasets by human annotation is costly and time-consuming.


Speech data synthesis~\cite{huang2023text} is a promising way to address data scarcity, enabling advancements in domain adaptation, recognition of rare names, numeric transcription, and low-resource languages~\cite{joshi2022simple,fazel2021synthasr,zevallos2022data}. Recent work leverages LLMs for text generation and multi-speaker TTS models for speech synthesis~\cite{cornell2024generating}. However, effective integration requires high-quality TTS models that produce naturalistic audio, as excessive synthesized data can degrade ASR performance on spontaneous and conversational speech.

Large amounts of untranscribed audio data are often available, but the lack of transcriptions limits their usability. A common approach is iterative pseudo-labeling, where unlabeled data is transcribed over multiple iterations as the acoustic model evolves~\cite{xu2020iterative}. The accuracy of pseudo-labels depends on the strength of the base ASR model, and its errors can propagate to the final model.

Another common approach for transcribing large audio datasets is combining multiple ASR models. Ensemble methods have been widely used to enhance ASR performance. The NIST recognizer output voting error reduction (ROVER) method integrates ASR outputs through a voting mechanism~\cite{fiscus1997post,schwenk2000combining}, while other approaches employ machine learning techniques for model fusion~\cite{utsuro2004empirical,matsushita2003evaluating}. Additionally, researchers have explored combining ASR models of various architectures at the hypothesis level to improve accuracy~\cite{wong2020combination,swietojanski2013revisiting}.

Few studies have explored combining multiple end-to-end (E2E) ASR models. It is often assumed that the extensive parameters of a single E2E ASR model provide enough flexibility to handle diverse speech domains, including noisy and varied speech styles. However, different E2E models exhibit varying accuracy across domains and accents. Hojo et al.~\cite{hojo2023combining} integrate acoustic information from multiple E2E ASR models with an external language model (LM) tailored to the target domain. Text-based methods for improving ASR transcriptions include LM rescoring~\cite{kolehmainen2023personalization} and neural LM-based error correction, which converts incorrect recognitions to ground truth sentences~\cite{zhang2019automatic,hrinchuk2020correction}. Additionally, sequence-to-sequence multimodal ASR error correction models have been proposed~\cite{lin2023multi,li2024crossmodal,dong2024cross}. However, ensemble and LM-based approaches require careful tuning to align with the error patterns of the underlying ASR model.


The rise of generative LLMs has sparked growing interest in using them as ASR correctors~\cite{nie2022prompt,koilakuntla2024leveraging,yang2023generative,ma2023can,pu2023multi}. Authors of ~\cite{chen2023hyporadise} leverage N-best lists generated by ASR and perform LoRA finetuning of LLM for generating best hypothesis. Instead of 1-best or N-best hypotheses, ~\cite{everson2024towards} utilize word confusion networks generated by the ASR system and perform ASR error correction demonstrating improved performance. Hu et al.~\cite{hu2024listen} further extend textual approaches by developing a multimodal model that incorporates discrete audio tokens as an additional input. Here, we consider a speechLLM  architecture~\cite{cui2025recentadvancesspeechlanguage}, a versatile recent variant of multimodal LLMs,  based on continuous audio embeddings. However, there is a lack of comparative studies evaluating the effectiveness of LLM-based generative approaches against multi-ASR ensemble methods for pseudo-label generation and ASR error correction.

\begin{figure*}[t]
 \centering
\centerline{\epsfig{figure=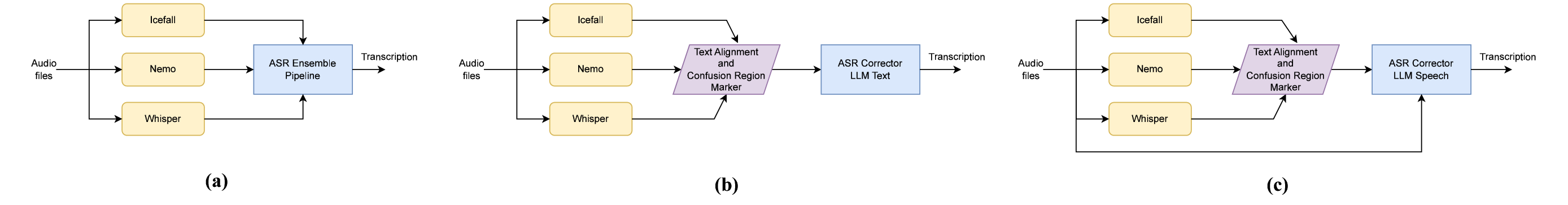,scale=0.38}}
\vspace{-.4cm}
\caption{Comparison of different approaches for generating pseudo labels, (a) Multi-ASR ensemble pipeline, (b) Multi-ASR textual LLM-based architecture, (c) Multi-ASR speechLLM-based architecture.}
\label{fig:three_approaches}
\vspace{-0.3cm}
\end{figure*}
In this work, we conduct a comparative study of three distinct approaches for generating pseudo-labels from untranscribed audio data, which are then used to train ASR models. First, we introduce a {\it multi-ASR ensemble pipeline}, which integrates three well-established large-scale end-to-end ASR models. We hypothesize that each ASR model brings unique perspectives to transcription generation, and by effectively combining them, we can produce more accurate hypotheses. Next, we propose a {\it multi-ASR textual LLM-based architecture}, where the prompt includes the confusion sets generated from the three ASR models.
While this method improves upon the ensemble approach by refining transcriptions, it disregards the acoustic information present in the original speech. Although acoustic or confidence scores from the ASR models can partially compensate for this limitation, incorporating the original speech signals into the correction process remains a key challenge. To address this, we propose a novel {\it multi-ASR speechLLM-based architecture}, which adopts a more comprehensive approach leveraging both textual hypotheses and acoustic evidence. This model is finetuned to learn from disagreements among the ASR ensemble. Finally, we validate that transcriptions generated by the strategies proposed achieve performance comparable to human annotation, when used for ASR training. Multi-ASR speechLLM-based error correction provides an effective way to utilize large-scale untranscribed audio data in semi-supervised ASR training.

The rest of the paper is organized as follows. In Section~\ref{sec:paseudo-labelling}, we describe the three approaches for generating pseudo labels. Section~\ref{sec:experiments} presents the database, experimental setup and results and we provide our conclusions in Section~\ref{sec:conclusion}.

\vspace{-0.2cm}
\section{Architectures for Pseudo-Labeling}\label{sec:paseudo-labelling}

We describe three architectures for the generation of pseudo-labels for untranscribed speech: (a) a multi-ASR ensemble pipeline, (b) multi-ASR with textual LLM postprocessing, and (c) multi-ASR with speechLLM postprocessing, as illustrated in Figure~\ref{fig:three_approaches}(a), ~\ref{fig:three_approaches}(b) and ~\ref{fig:three_approaches}(c), respectively. All approaches integrate three ASR models— Custom Icefall model, Nemo Parakeet, and OpenAI Whisper. The conventional ensemble pipeline applies a complex cascade of rule-based processing to derive final transcriptions. In contrast, the textual LLM-based approach simplifies this by directly structuring ASR outputs for LLM finetuning. The speechLLM-based approach further enhances this by incorporating both textual and speech inputs, leveraging audio evidence for more accurate predictions.



\vspace{-0.2cm}
\subsection{Multi-ASR Ensemble Pipeline}
\label{sec:ensemble}

\begin{figure}[t]
\centering
\centerline{\epsfig{figure=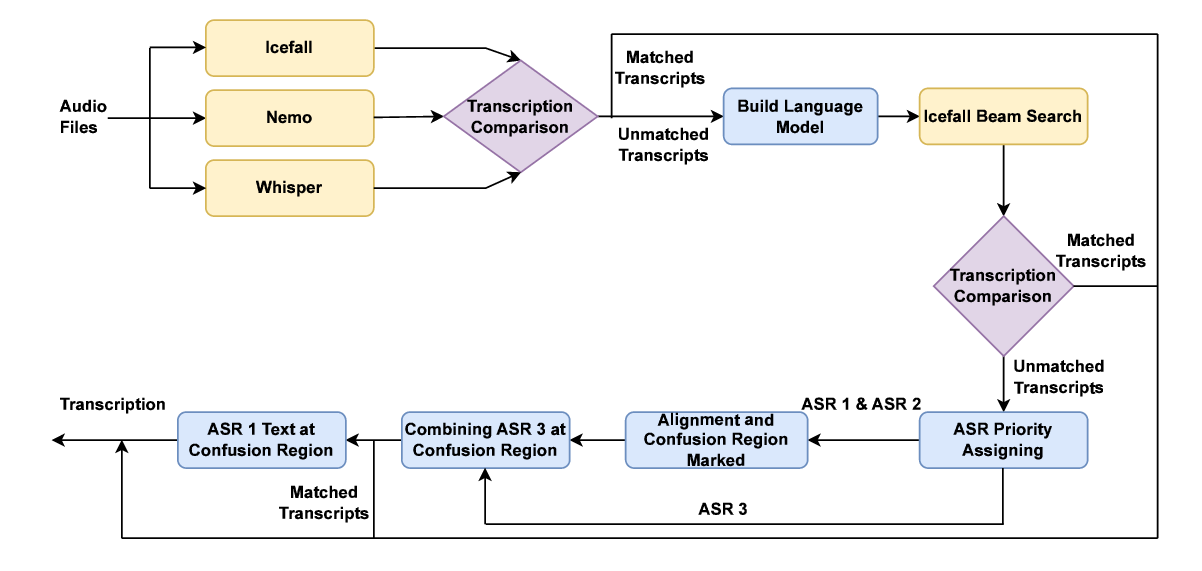,scale=0.44}}
{\ninept \caption{ Multi-ASR ensemble pipeline for pseudo-labeling.}
\label{fig:framework}}
\vspace{-0.6cm}
\end{figure}

In this framework, we perform the fusion of outputs from all three recognizers and use the consensus of a majority at the word level to generate a final transcription. 
As shown in Figure~\ref{fig:three_approaches}(a), 
we pass the audio concurrently through the three ASR models, namely, Icefall\footnote{https://github.com/k2-fsa/icefall}, Nemo Parakeet\footnote{https://rb.gy/a9y6c3}, and OpenAI Whisper\footnote{https://huggingface.co/openai/whisper-large-v3}. As shown in Figure~\ref{fig:framework}, the multi-ASR ensemble pipeline uses one-pass decoding for Nemo and Whisper, while Icefall follows a two-pass strategy.  In the first pass, we decode the audio using a greedy strategy, and the character error rate (CER) is computed between each pair of ASR outputs. Utterances with zero CER (perfect matches) are excluded from further processing. Non-matching transcriptions proceed to a second decoding pass, where we use an LM. We train the LM using all the unmatched transcripts generated during the first-pass decoding, which is an offline process. The second decoding pass, using Icefall's fast beam search, refines uncertain outputs. A final CER-based comparison aligns Icefall’s second-pass outputs with Nemo and Whisper transcriptions.



Based on the CERs of the pairs of mismatched utterances, we rank the three ASR models as ASR-1, ASR-2, and ASR-3, such that ASR model with the lowest average CER is designated as ASR-1. A textual alignment is performed between ASR-1 and ASR-2, and confusion regions are marked, thus highlighting the region of uncertainty between the two transcripts. Transcription from ASR-3 is considered to resolve this uncertainty between ASR-1 and ASR-2. An alignment is performed between ASR-3 and the confusion regions of ASR-1 and ASR-2. If ASR-3 agrees with any of the ASR-1 or ASR-2 transcripts, the uncertainty is resolved, and that region is marked as confident. For the remaining uncertain regions, the transcript from ASR-1 is used to generate the pseudo-labels.

\vspace{-0.2cm}
\subsection{Multi-ASR Textual LLM-based Architecture}\label{sec:textual-LLM}
In this architecture, we utilize LLMs that have strong capabilities of text understanding, contextual reasoning and text generation. We use  Llama 3.2 1B instruct model~\cite{grattafiori2024llama3herdmodels} as the textual LLM to refine uncertain ASR outputs by its learned language and world knowledge, and ability to follow task specific instructions. However, to effectively handle ASR ensemble outputs particularly in the form of textual confusion networks, finetuning is necessary for task specific adaptation.


As shown in Figure~\ref{fig:three_approaches}(b), we perform text alignment between the hypotheses generated from the three ASR models and construct confusion networks in textual form that highlight areas of disagreement and uncertainty. For finetuning the LLM, we prepare the instruction tuning data. The instructions assign a role to the LLM, describes the task and includes the confusion set generated from the text alignment module.

Instead of relying on manually engineered rules, extensively optimized heuristics and a pipeline of post-processing modules, the textual LLM learns how to weigh different hypotheses, correct ASR errors and produce more accurate transcriptions in an end-to-end manner.
\vspace{-0.2cm}
\subsection{Multi-ASR speechLLM-based Architecture}~\label{sec:speech_llm}
\vspace{-0.4cm}

While the confusion networks as described in Section~\ref{sec:textual-LLM} provide valuable textual information to LLMs for correcting the transcriptions, the acoustic information present in the original speech is ignored. With speechLLMs, we can create a more holistic system that considers the textual hypotheses along with the underlying acoustic evidence.

In the proposed speechLLM-based framework, we initially derive the hypotheses from three ASR models and create the textual confusion network as described in Section~\ref{sec:textual-LLM}. We prepare the training data, which are comprised of triplets that include a speech waveform, an instruction with transcription containing confusion sets, and a ground truth response as target output. We create this training data by adding the original audio to the finetuning data created for the textual LLM as follows.
\begin{Verbatim}[commandchars=\\\{\},
                 breaklines=true,
                 breakanywhere=true,
                 breaksymbolleft=,
                 breaksymbolright=]
    \textbf{audio_path}: "sample_xxx.wav"
    \textbf{instruction}: "Use the text provided and correct the mistakes made by ASR. For better reliability 3 ASRs are used for transcription and the low confidence regions are marked as confusion regions within different brackets ` {all}|<>|[] right  {two}|<too>|[too] bye`"
    \textbf{response}: "all right you too bye"
\end{Verbatim}

Although there exist several multimodal LLM architectures, we select a model combining a speech encoder and a textual LLM, linked by a lightweight adapter for modality alignment~\cite{cui2025recentadvancesspeechlanguage,das2024speechverselargescalegeneralizableaudio}. This data-efficient approach requires minimal task-specific supervision since the foundational models are pretrained on large-scale audio and text data. However, parameter-efficient finetuning is essential for modality alignment, task adaptation, and domain adaptation. The modality alignment through the adapter ensures that the speech encoder output is effectively mapped into the textual LLM's input representation space, allowing the model to interpret and utilize acoustic information along with ASR hypotheses and instructions. The task adaptation through the textual LLM ensures that it correctly interprets its multimodal input and generates the outputs aligned with the task's requirements. Additionally, finetuning of speech encoder and textual LLM achieves domain adaptation ensuring the model handles the acoustic conditions and linguistic characteristics relevant to the product setups and use cases. By considering these three objectives in a unified manner and finetuning all components jointly, we achieve performance levels surpassing the textual LLM-based method.

\vspace{-0.2cm}
\section{Experiments}\label{sec:experiments}
\subsection{Datasets used}\label{dataset}
\vspace{-0.1cm}
To effectively compare the presented strategies for generating pseudo-labels, we use diverse datasets that include conversational, telephony and read-speech data in different domains from both  public and in-house sources. Table~\ref{tab:databse_details} summarizes the details of ASR training, LLM finetuning and test corpora used in our experiments. For the training sets in Table~\ref{tab:databse_details}, we generate pseudo-labels using different approaches, which are then used to train corresponding ASR models. LLM finetuning data, comprising 27K training samples, is used for finetuning both text and speechLLMs. Since the foundation models have already been trained on vast amounts of audio and text, our focus is on constructing an efficient task-specific dataset that provides high-quality supervision. We employ different test sets to evaluate the WER between ground-truth labels and pseudo-labels generated by the different architectures. The test sets include three in-house call center datasets, DefinedAI data\footnote{Website: https://www.defined.ai}, multi-domain data from Wow AI\footnote{https://wow-ai.com/data.html} along with open source datasets. Since these datasets are not open-source, we also ensure that they are not seen by any external open-source ASR or language models, thus avoiding data contamination. There is no overlap between the DefinedAI training, LLM finetuning, and test data.



\begin{table}[t]
\caption{\footnotesize{Duration and domain information for different training and test sets used in the experiments.}}
\label{tab:databse_details} 
\renewcommand{\arraystretch}{1.1}
\resizebox{0.48\textwidth}{!}{
\begin{tabular}{|c c c |}
\hline
\multicolumn{1}{|c|}{Dataset} & \multicolumn{1}{|c|}{Duration (hours)} & \multicolumn{1}{|c|}{Domains}\\
\hline
\hline
\multicolumn{3}{|c|}{Train data}\\
\hline
\hline
\multicolumn{1}{|c|}{Librispeech} & \multicolumn{1}{|c|}{960} & \multicolumn{1}{|c|}{Audiobook }\\
\hline
\multicolumn{1}{|c|}{DefinedAI} & \multicolumn{1}{|c|}{1410} & \multicolumn{1}{|c|}{Banking, Insurance, Retail, Telco}\\
\hline
\hline
\multicolumn{3}{|c|}{LLM finetuning data}\\
\hline
\hline
\multicolumn{1}{|c|}{DefinedAI} & \multicolumn{1}{|c|}{289} & \multicolumn{1}{|c|}{Banking, Insurance, Retail, Telco}\\
\hline
\hline
\multicolumn{3}{|c|}{Test data}\\
\hline
\hline
\multicolumn{1}{|c|}{Wow} & \multicolumn{1}{|c|}{18.21} & \multicolumn{1}{|c|}{Autoinsurance, Automotive, Medicare}\\
\hline
\multicolumn{1}{|c|}{Wow} & \multicolumn{1}{|c|}{18.21} & \multicolumn{1}{|c|}{Medical, Home Service, Customer Service}\\
\hline
\multicolumn{1}{|c|}{Call center 1} & \multicolumn{1}{|c|}{2.00} & \multicolumn{1}{|c|}{Medical}\\
\hline
\multicolumn{1}{|c|}{Call center 2} & \multicolumn{1}{|c|}{7.62} & \multicolumn{1}{|c|}{Telco}\\
\hline
\multicolumn{1}{|c|}{Call center 3} & \multicolumn{1}{|c|}{12.29} & \multicolumn{1}{|c|}{Telco}\\
\hline
\multicolumn{1}{|c|}{Gigaspeech~\cite{chen2021gigaspeech}} & \multicolumn{1}{|c|}{36.92} & \multicolumn{1}{|c|}{Multi-domain}\\
\hline
\multicolumn{1}{|c|}{Librispeech test-clean} & \multicolumn{1}{|c|}{5} & \multicolumn{1}{|c|}{Audiobook}\\
\hline
\multicolumn{1}{|c|}{Librispeech test-other} & \multicolumn{1}{|c|}{5} & \multicolumn{1}{|c|}{Audiobook}\\
\hline
\multicolumn{1}{|c|}{DefinedAI banking} & \multicolumn{1}{|c|}{30.27} & \multicolumn{1}{|c|}{Banking}\\
\hline
\multicolumn{1}{|c|}{DefinedAI insurance} & \multicolumn{1}{|c|}{39.70} & \multicolumn{1}{|c|}{Insurance}\\
\hline
\multicolumn{1}{|c|}{DefinedAI retail} & \multicolumn{1}{|c|}{33.02} & \multicolumn{1}{|c|}{Retail}\\
\hline
\multicolumn{1}{|c|}{DefinedAI telco} & \multicolumn{1}{|c|}{42.76} & \multicolumn{1}{|c|}{Telco}\\
\hline
\end{tabular}
}
\vspace{-.4cm}
\end{table}
\vspace{-0.2cm}

\subsection{Experimental setup}

We compare the various pseudo-labeling methods by computing transcription accuracy for various test and training sets, and finally training Icefall models with the training sets and their automatic transcripts. 

\vspace{-0.2cm}
\subsubsection{Multi-ASR ensemble pipeline}\label{sec:exp-ensemble}
The Icefall model that is part of the ASR ensemble is trained in-house using the standard Zipformer recipe from the Icefall toolkit with 65 million parameters. Training used DefinedAI conversational data combined with in-house real and simulated call center data, totaling 6,600 hours of training data. The pretrained Nvidia Parakeet RNNT model, with 1.1 billion parameters, was trained on 64,000 hours of data, including both public and proprietary datasets. The pretrained OpenAI Whisper-large-v3 model, with 1.5 billion parameters, was trained on a 680,000 hours of multilingual audio data, incorporating both weakly labeled and pseudo-labeled data. 

We perform decoding in parallel using 3 NVIDIA A10 GPUs in the first pass and single NVIDIA A10 GPU in the second pass. In second pass, the Icefall K2 framework is used to build the language model, and the decoding method used is fast beam search n-best LG, where beam search is used to efficiently generate the top n-best hypotheses guided by an LM to improve the transcription quality. The number of paths $n$ is set to 200. After second-pass decoding, textual alignment is performed based on Levenstein distance as a heuristic. 
\vspace{-0.2cm}
\subsubsection{Multi-ASR Textual LLM-based Architecture}\label{sec:exp_textual_llm}

\begin{table}[t]
\caption{\footnotesize{WER (\%) of pseudo-labels generated from different individual ASR models as well as with multi-ASR ensemble pipeline, textual LLM postprocessing, and speechLLM-based postprocessing, with respect to the ground-truth transcriptions.}}
\label{tab:ensemble_results} 
\renewcommand{\arraystretch}{1.2}
\vspace{-0.2cm}
\scalebox{0.67}{
\begin{tabular}{|c c c c c c c|}
\hline
\multicolumn{1}{|c|}{Dataset} & \multicolumn{1}{|c|}{Icefall} & \multicolumn{1}{|c|}{Nemo} & \multicolumn{1}{|c|}{Whisper} & \multicolumn{1}{|c|}{Multi-ASR}  & \multicolumn{1}{|c|}{Textual} & \multicolumn{1}{|c|}{Speech}\\
\multicolumn{1}{|c|}{} & \multicolumn{1}{|c|}{} & \multicolumn{1}{|c|}{} & \multicolumn{1}{|c|}{} & \multicolumn{1}{|c|}{ensemble}  & \multicolumn{1}{|c|}{ LLM} & \multicolumn{1}{|c|}{LLM}\\
\hline
\hline
\multicolumn{7}{|c|}{Train data} \\
\hline
\hline
\multicolumn{1}{|c|}{Defined} & \multicolumn{1}{|c|}{15.63} & \multicolumn{1}{|c|}{14.54} & \multicolumn{1}{|c|}{19.36} & \multicolumn{1}{|c|}{14.36}  & \multicolumn{1}{|c|}{11.60}   & \multicolumn{1}{|c|}{\textbf{9.30}}\\
\hline
\multicolumn{1}{|c|}{Librispeech} & \multicolumn{1}{|c|}{9.37} & \multicolumn{1}{|c|}{\textbf{1.12}} & \multicolumn{1}{|c|}{3.30} & \multicolumn{1}{|c|}{1.94}  & \multicolumn{1}{|c|}{2.83}   & \multicolumn{1}{|c|}{1.95}\\
\hline
\hline
\multicolumn{7}{|c|}{Test data} \\
\hline
\hline
\multicolumn{1}{|c|}{Call center 1} & \multicolumn{1}{|c|}{12.04} & \multicolumn{1}{|c|}{14.97} & \multicolumn{1}{|c|}{19.11} & \multicolumn{1}{|c|}{12.92}  & \multicolumn{1}{|c|}{11.39}   & \multicolumn{1}{|c|}{\textbf{9.85}}\\
\hline
\multicolumn{1}{|c|}{Call center 2} & \multicolumn{1}{|c|}{14.47} & \multicolumn{1}{|c|}{17.41} & \multicolumn{1}{|c|}{21.11} & \multicolumn{1}{|c|}{14.76}  & \multicolumn{1}{|c|}{14.60}   & \multicolumn{1}{|c|}{\textbf{13.57}}\\
\hline
\multicolumn{1}{|c|}{Call center 3} & \multicolumn{1}{|c|}{15.32} & \multicolumn{1}{|c|}{18.10} & \multicolumn{1}{|c|}{23.69} & \multicolumn{1}{|c|}{15.52}  & \multicolumn{1}{|c|}{15.29}   & \multicolumn{1}{|c|}{\textbf{14.59}}\\
\hline
\multicolumn{1}{|c|}{Wow} & \multicolumn{1}{|c|}{13.29} & \multicolumn{1}{|c|}{17.13} & \multicolumn{1}{|c|}{17.70} & \multicolumn{1}{|c|}{12.97}  & \multicolumn{1}{|c|}{\textbf{12.49}}   & \multicolumn{1}{|c|}{12.57}\\
\hline
\multicolumn{1}{|c|}{Gigaspeech} & \multicolumn{1}{|c|}{15.89} & \multicolumn{1}{|c|}{13.32} & \multicolumn{1}{|c|}{16.13} & \multicolumn{1}{|c|}{12.99}  & \multicolumn{1}{|c|}{12.10}   & \multicolumn{1}{|c|}{\textbf{11.75}}\\
\hline
\multicolumn{1}{|c|}{Librispeech test-clean} & \multicolumn{1}{|c|}{8.15} & \multicolumn{1}{|c|}{\textbf{1.74}} & \multicolumn{1}{|c|}{3.30} & \multicolumn{1}{|c|}{2.22}  & \multicolumn{1}{|c|}{2.96}   & \multicolumn{1}{|c|}{2.26}\\
\hline
\multicolumn{1}{|c|}{Librispeech test-other} & \multicolumn{1}{|c|}{15.11} & \multicolumn{1}{|c|}{\textbf{3.08}} & \multicolumn{1}{|c|}{5.66} & \multicolumn{1}{|c|}{4.06}  & \multicolumn{1}{|c|}{5.21}   & \multicolumn{1}{|c|}{4.50}\\
\hline
\multicolumn{1}{|c|}{DefinedAI banking} & \multicolumn{1}{|c|}{13.38} & \multicolumn{1}{|c|}{16.47} & \multicolumn{1}{|c|}{19.88} & \multicolumn{1}{|c|}{13.85}  & \multicolumn{1}{|c|}{11.67}   & \multicolumn{1}{|c|}{\textbf{10.45}}\\
\hline
\multicolumn{1}{|c|}{DefinedAI insurance} & \multicolumn{1}{|c|}{13.53} & \multicolumn{1}{|c|}{15.70} & \multicolumn{1}{|c|}{19.07} & \multicolumn{1}{|c|}{13.77}  & \multicolumn{1}{|c|}{11.18}   & \multicolumn{1}{|c|}{\textbf{9.45}}\\
\hline
\multicolumn{1}{|c|}{DefinedAI retail} & \multicolumn{1}{|c|}{15.39} & \multicolumn{1}{|c|}{17.45} & \multicolumn{1}{|c|}{20.38} & \multicolumn{1}{|c|}{15.35}  & \multicolumn{1}{|c|}{12.97}   & \multicolumn{1}{|c|}{\textbf{11.79}}\\
\hline
\multicolumn{1}{|c|}{DefinedAI telco} & \multicolumn{1}{|c|}{13.71} & \multicolumn{1}{|c|}{15.58} & \multicolumn{1}{|c|}{18.83} & \multicolumn{1}{|c|}{13.75}  & \multicolumn{1}{|c|}{11.25}   & \multicolumn{1}{|c|}{\textbf{10.21}}\\
\hline
\end{tabular}
}
\vspace{-.9cm}
\end{table}
For postprocessing using a textual LLM, we use Llama 3.2 1B instruct as the LLM.  The software ecosystem for this experiment is based primarily on the Huggingface framework. Model development and experiments were conducted using four NVIDIA
A10G GPUs with 24GB of memory each. We finetune the textual LLM using the trainer class with QLoRA (4-bit) applied to all its linear layers. The LoRA rank was selected as 32 with $\alpha = 128$; we used a dropout rate of 0.05. The batch size was set to 4 per GPU, with gradient accumulation of 8 steps, resulting in an effective batch size of 128. The learning rate was managed using a cosine scheduler over 10 to 15 epochs, with a maximum learning rate of $2\times 10^{-4}$. We implement a linear warm-up for the first 20\% of the total number of iterations. For optimization, we employ the AdamW (weighted Adam) optimizer. We use its default parameters including $\beta_1 = 0.9$, $\beta_2 = 0.999$, and $\epsilon = 10^{-8}$, without applying any weight decay. For inference, we use the greedy search with temperature 0. The number of new tokens is limited to 512. 
\vspace{-0.2cm}
\subsubsection{Multi-ASR SpeechLLM-based Architecture}
For speechLLM postprocessing we use the Qwen2-Audio speech/audio foundation model~\cite{chu2024qwen2audiotechnicalreport}, which has been natively integrated into the HuggingFace ecosystem. The finetuning and inference setup for this model is same as for the textual LLM, as described in Section~\ref{sec:exp_textual_llm}.

\vspace{-0.2cm}
\subsubsection{ASR model training with pseudo labels}
We train two sets of Icefall ASR models using the generated pseudo-labels, one using Librispeech training data, the other using DefinedAI training data, as described in Table~\ref{tab:databse_details}. For each training set, we train versions based on ground truth transcription as well as pseudo-transcriptions from the multi-ASR ensemble, textual LLM postprocessing, and speechLLM postprocessing pipelines. All the Icefall ASR models mentioned here are trained using the standard Zipformer recipe from the Icefall toolkit and the training setup is same as mentioned in~\ref{sec:exp-ensemble}.

\vspace{-0.2cm}
\subsection{Results}
\vspace{-0.1cm}
We first evaluate the accuracy of the various pseudo-labeling methods employing the three different frameworks as well as each individual ASR model, on all training and test sets listed in Table~\ref{tab:databse_details}.
As a performance metric, we use word error rate (WER) between the ground truth transcription and the machine-generated transcription, as shown in Table~\ref{tab:ensemble_results}.  We observe that the top-performing ASR varies depending on the datasets, making it challenging to select a single ASR system for generating pseudo-labels. The multi-ASR ensemble achieves a better/comparable WER to the best-performing individual ASR, highlighting the approach's ability to effectively combine the Icefall, Nemo, and Whisper models. The textual LLM outperforms the ensemble approach and all individual ASRs, except on the Librispeech datasets. This can be attributed to the fact that the Nemo model has seen Librispeech data during training. The improvement is further enhanced with the speechLLM-based corrector. The fusion of audio with the textual confusion network allows the speechLLM to better disambiguate among the generated choices, compared to the textual LLM. For both types of LLM, however, the improvement is greater for the DefinedAI datasets, due to the domain alignment with the finetuning data, demonstrating the importance of domain adaptation. At the same time, we see improvements for other domains, showing the robustness of LLM-based postprocessing.

Table~\ref{tab:libri_iefall} shows the WERs for the Icefall ASR models trained with ground truth transcription and the pseudo-transcriptions generated by different approaches on either Librispeech or DefinedAI data. The evaluation uses matched test sets for each of the training sets. For Librispeech test-clean data, the model based on speechLLM transcriptions achieved 3.22\% WER compared to 3.40\% using ground truth transcriptions. For test-other, the speechLLM-based result is 8.34\% while the one based on just the multi-ASR ensemble is 8.44\%.
For DefinedAI-trained models, speechLLM-generated transcriptions outperform human-labeled ground truth, highlighting possibly better-than-human (but not easily measurable) transcription accuracy. Additionally, ASR using textual LLM postprocessing surpasses the multi-ASR ensemble approach.\footnote{We also performed experiments with the LLM finetuning data {\em added} to the training of the baseline ASR methods, which reduced WER by only 0.1\% to 0.2\% absolute. This confirms that the additional data used in LLM learning is {\em not} the primary source of the observed gains.}

In all cases, the results show that the proposed speechLLM-based transcription method generates transcripts close to ground truth, outperforming all other pseudo-labeling approaches.

\begin{table}[t]
\caption{\footnotesize{WER (\%) of ASR models trained using ground truth text, a single ASR system output (Icefall), or transcriptions from the proposed ensemble pipelines, for both Librispeech and DefinedAI data.}}
\vspace{-0.2cm}
\label{tab:libri_iefall} 
\renewcommand{\arraystretch}{1.2}
\resizebox{0.47\textwidth}{!}{
\begin{tabular}{|c c c c c|}
\hline
\multicolumn{1}{|c|}{Training data$\rightarrow$} & \multicolumn{4}{|c|}{Librispeech training}\\
\hline
\multicolumn{1}{|c|}{Test Data$\downarrow$} & \multicolumn{1}{|c|}{Ground truth} & \multicolumn{1}{|c|}{Multi-ASR ensemble} & \multicolumn{1}{|c|}{Textual LLM}  & \multicolumn{1}{|c|}{SpeechLLM}\\
\hline
\hline
\multicolumn{1}{|c|}{Librispeech test-clean} & \multicolumn{1}{|c|}{3.40} & \multicolumn{1}{|c|}{3.38} & \multicolumn{1}{|c|}{3.87}  & \multicolumn{1}{|c|}{3.22}\\
\hline
\multicolumn{1}{|c|}{Librispeech test-other} & \multicolumn{1}{|c|}{8.32} & \multicolumn{1}{|c|}{8.44} & \multicolumn{1}{|c|}{8.93}  & \multicolumn{1}{|c|}{8.34}\\
\hline
\hline
\multicolumn{1}{|c|}{Training data$\rightarrow$} & \multicolumn{4}{|c|}{DefinedAI training}\\
\hline
\hline
\multicolumn{1}{|c|}{DefinedAI banking} & \multicolumn{1}{|c|}{9.0} & \multicolumn{1}{|c|}{10.3}   & \multicolumn{1}{|c|}{9.1}  & \multicolumn{1}{|c|}{8.7}\\
\hline
\multicolumn{1}{|c|}{DefinedAI insurance} & \multicolumn{1}{|c|}{8.2} & \multicolumn{1}{|c|}{9.5}   & \multicolumn{1}{|c|}{8.4}  & \multicolumn{1}{|c|}{8.0}\\
\hline
\multicolumn{1}{|c|}{DefinedAI retail} & \multicolumn{1}{|c|}{11.2} & \multicolumn{1}{|c|}{12.3} & \multicolumn{1}{|c|}{11.4}  & \multicolumn{1}{|c|}{10.9}\\
\hline
\multicolumn{1}{|c|}{DefinedAI telco} & \multicolumn{1}{|c|}{8.8} & \multicolumn{1}{|c|}{9.9}  & \multicolumn{1}{|c|}{8.9}  & \multicolumn{1}{|c|}{8.6}\\
\hline
\end{tabular}
}
\vspace{-.4cm}
\end{table}

\vspace{-.2cm}
\section{Conclusions}\label{sec:conclusion}
\vspace{-.1cm}
We have explored the use of textual and speechLLMs to improve ASR post-processing for pseudo-labeling, moving beyond a traditional cascaded method. The introduction of LLMs enable a unified framework, where a textual LLM refines ASR outputs based on confusion networks and a speechLLM further consults the audio data for improved disambiguation decisions, in a single step. After casting the post-processing as an instruction-following task we have finetuned the base models in a parameter- and data-efficient manner. Thorough evaluations on multiple datasets covering diverse domains and acoustic conditions we demonstrate that textual LLMs significantly outperform a conventional multi-ASR approach and speechLLMs provide further gains by jointly using speech and textual inputs. Our findings highlight the effectiveness of LLM-driven approaches for pseudo-labeling, and hence, for semi-supervised ASR training and adaptation. In short, by unifying pseudo-labeling into a single-stage instruction-following framework, we significantly simplify the process while avoiding error propagation, information loss and disjoint optimization, resulting in improved pseudo-labels for semi-supervised ASR training.


\newpage
\section{Acknowledgments}
We would like to express our sincere gratitude to Aravind Ganapathiraju for his invaluable guidance. We also extend our thanks to Rohit Mishra and Stephen L for their support.

\bibliographystyle{IEEEtran}
\bibliography{mybib}


\end{document}